\documentstyle[aps,prd,epsf]{revtex}
\newcommand{\bea}{\begin{eqnarray}}  \newcommand{\eea}{\end{eqnarray}}
\newcommand{\beqa}{\begin{eqnarray}}  \newcommand{\eeqa}{\end{eqnarray}}
\newcommand{\beq}{\begin{equation}}  \newcommand{\eeq}{\end{equation}}
  
\newcommand{\lmk}{\left(}  \newcommand{\rmk}{\right)}
  
\newcommand{\lkk}{\left[}  \newcommand{\rkk}{\right]}

\newcommand{\vect}[1]{\mbox{\boldmath${#1}$}}
\newcommand{\thin}{{\rm thin}}
\newcommand{\app}{{\rm app}}
\newcommand{\fbar}{\overline{f}}
\newcommand{\gbar}{\overline{g}}

\begin{document}
\draft
%
%
\vspace*{-20mm}
\leftline{\epsfbox{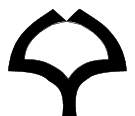}}
\vspace*{-10mm}
{\baselineskip-4pt
\font\yitp=cmmib10 scaled\magstep2
\font\elevenmib=cmmib10 scaled\magstep1  \skewchar\elevenmib='177
\leftline{\baselineskip20pt
\hspace{12mm} 
\vbox to0pt
   { {\yitp\hbox{Osaka \hspace{1.5mm} University} }
     {\large\sl\hbox{{Theoretical Astrophysics}} }\vss}}

%
%
{\baselineskip0pt
\rightline{\large\baselineskip14pt\rm\vbox
        to20pt{\hbox{February 2003}
               \hbox{OU-TAP-187}
\vss}}
}
\vskip20mm

\begin{center}
{\Large\bf   Cosmic Inversion II\\
~~\\
---An iterative method for reproducing 
the primordial spectrum from the CMB data ---}
\end{center}
\vspace{3mm}

\begin{center}
{\large
 Makoto Matsumiya,\footnote{E-mail:matumiya@vega.ess.sci.osaka-u.ac.jp}
  Misao Sasaki\footnote{E-mail:misao@vega.ess.sci.osaka-u.ac.jp}
  and \,Jun'ichi Yokoyama\footnote{E-mail:yokoyama@vega.ess.sci.osaka-u.ac.jp}}
\vskip 1mm
\sl{Department of Earth and Space Science,
Graduate School of Science,\\
Osaka University, Toyonaka 560-0043, Japan}
\end{center}

\begin{abstract}
Pursuing the original idea proposed in our previous paper (Paper I),
we improve the method to determine the shape
of the initial curvature perturbation
spectrum $P(k)$ from the CMB data.
The thickness of the last scattering surface (LSS)
 and the integrated Sachs-Wolfe
(ISW) effect, which we neglect in Paper I, are taken into account and
 an iterative method is newly developed.
The new method can reproduce
the primordial power spectra with a high accuracy, given the
correct values of the cosmological parameters. Conversely,
there appear spurious peaks and dips in the reconstructed
power spectrum if we use the cosmological parameters
slightly different from the true values, while there
appear regions of negative $P(k)$ in some cases
if we use substantially different values.
In other words, the tacit assumption that the cosmological parameters
 can be determined for an assumed initial spectrum is verified by 
our reconstruction method. In addition, it could be a new tool
to constrain the cosmological parameters without recourse to
models of the primordial power spectra.
\end{abstract}


\section{Introduction}
Cosmic microwave background (CMB) is now playing a central
role in the precision cosmology and enables us to extract wealth of
information on the cosmological parameters and the primordial universe.
The past CMB experiments suggest that our universe is consistent with
spatially flat $\Lambda$CDM universe with scale-invariant initial
spectrum \cite{FCDM}, which is predicted by conventional slow-roll
inflation models \cite{PER}.
Nevertheless these results are not too restrictive and
other cosmological models are still possible under the present
accuracy of the observational data.
In the near future, we will attain unprecedentedly precise data by MAP
\cite{MAP} and
Planck \cite{Planck} experiments. Then, we will be able to
 discuss the results much more
quantitatively than we can now.

In particular, on the analysis of the CMB anisotropy, it is more
desirable to deal with the initial spectrum of perturbation 
as an arbitrary
function, although  most of the past analyses of observational data
assumed a power-law spectrum.
On the theoretical side, a variety of generation mechanisms of broken
scale-invariant spectra have been proposed, even in the context
of inflationary cosmology \cite{BSI}.  For this reason or others,
there have been some attempts to
reconstruct the initial spectrum from the data \cite{SOU,BER,TEG}.
But some of them assumed that the initial spectrum is a
piecewise power-law function or parametrised 
as a broken scale-invariant function \cite{SOU},
while others took only the Sachs-Wolfe (SW) effect \cite{SW} into account
\cite{BER}.

In our previous paper (hereafter Paper~I) \cite{MAT},
we built up a basic framework to reconstruct the primordial spectrum of
curvature perturbations, $P(k)$, out of CMB anisotropy data by solving a
differential equation for $P(k)$, in which not only the SW effect but also
the Doppler effect are taken into account.  In Paper~I, however, we
considered only a simplified and rather unrealistic
cosmological model for illustration
without any degrees of freedom of cosmological parameters.
Furthermore, we
neglected  the thickness of the last scattering surface (LSS), which
caused considerable errors in the angular power spectrum $C_{\ell}$.

In this paper, we improve the method to a more accurate one which can
be applied to realistic models.
 The width of the LSS and Integrated
Sachs-Wolfe term are newly included. We also investigate
 the dependence on the cosmological parameters.
Our inversion method assumes that the true cosmological parameters are
known. In reality, this is not the case. In fact, the
standard argument is to use the CMB data to determine the
cosmological parameters, with some assumptions on the initial power
spectrum. However, we cannot apply this argument to our case,
 since we treat the initial spectrum as an arbitrary function.
Thus, it is important to study the effect on the
reconstructed spectrum as we vary the cosmological parameters.
Since our main purpose here is to complete solving the 
inverse problem for realistic models, we assume an ideal situation where
the observational error does not exist. 

This paper is organized as follows. In Sec.~II, we describe our inversion
method in some detail, which is an improved version of the one presented
in Paper~I. In Sec.~III, we apply our method to some model spectra,
assuming
the true cosmological parameters are known.
We find that the new method can reproduce the primordial spectrum to
quite accurately. In Sec.~IV, we investigate the dependence of the
reconstructed
spectrum on the cosmological parameters. 
We find the cosmological parameters
can be constrained severely
without any assumption on the primordial spectrum.
We discuss the implications of our method in Sec.~V.

\section{Inversion Formula}

First we write down basic equations. We use the same notation as Paper~I.
Working in the Fourier space and denoting the direction cosine between
the wavenumber vector $\vect{k}$ and the momentum vector of the photon
by $\mu$, an integral form of
the Boltzmann equation for the temperature anisotropy,
$\Theta(\eta,\mu,k)$, is given in the Newton gauge by \cite{HS}
\begin{equation}
\Theta(\eta_0,k,\mu)+\Psi(\eta_0,k)=\int^{\eta_0}_0 \,
\left\{\lmk\Theta_0+\Psi-i\mu V_b\rmk
  {\cal V}(\eta)+(\dot{\Psi}-\dot{\Phi})e^{-\tau(\eta)}\right\}
                        e^{ik\mu(\eta-\eta_0)}d \eta. \label{Boltzmann}
\end{equation}
Here, an overdot denotes a derivative with
respect to the conformal time $\eta$, and $\eta_0$ is its present value,
$\Psi$ and $\Phi$ are the Newtonian potential
and spatial curvature perturbation on the Newton slices, respectively
 \cite{KS}, and $\Theta_0$ is the monopole component
of the multipole expansion,
\begin{equation}
\Theta(\eta,k,\mu)=\sum_{\ell} (-i)^{\ell} \Theta_{\ell}(\eta,k)
P_{\ell}(\mu).
\end{equation}
The function ${\cal V}(\eta)$ is called the visibility
function and $\tau(\eta)$ is the opacity, which are given by
\begin{eqnarray}
{\cal V}(\eta) =  \dot{\tau}(\eta)e^{-\tau(\eta)}\, ,
\quad
\tau(\eta) =  \int^{\eta_0}_{\eta} \dot{\tau}(\eta^{\prime})
                 d \eta^{\prime}, \quad \dot{\tau}=ax_e n_e\sigma_T\,,
\end{eqnarray}
where $a$ is the cosmic scale factor, $x_en_e$ is the free
electron density, and $\sigma_T$ is the Thomson cross section.
Note that we have neglected a term due to anisotropic stress in the
integrand of Eq.~(\ref{Boltzmann}).
As we will see below, however, this causes no problem in our
inversion method. Also, we note that $\Psi= -\Phi$ apart from the
effect of small anisotropic stress due to photons and neutrinos.

Under the thin LSS approximation adopted in Paper~I, these functions are
approximated by
\beq
  {\cal V}(\eta)=\delta(\eta-\eta_*),\quad\quad
e^{-\tau(\eta)}=\theta(\eta-\eta_*),
\eeq
respectively \cite{HS}, with $\eta_*$ being the decoupling epoch when
the visibility function is maximum.
Hence, if we neglect the ISW term,
$\Theta_{\ell}(\eta_0,k)$ is given by
\beq
\Theta_{\ell}^{\thin}(\eta_0,k)  =  (\Theta_0+\Psi)(\eta_{*},k)(2\ell+1)
j_{\ell}(kd)
+\Theta_{1}(\eta_*,k)(2\ell+1)j^{\prime}_{\ell}(kd), \label{thin}
\eeq
where $d=\eta_0-\eta_*$ is the conformal distance to the LSS.
Assuming primordial fluctuation is adiabatic, both
$(\Theta_0+\Psi)(\eta_{*},k)$ and $\Theta_{1}(\eta_*,k)$ can be
expressed as
\begin{eqnarray}
(\Theta_0+\Psi)(\eta_*,k) & = & f(\eta_*,k)\Phi(0,k),
\nonumber\\
\Theta_1(\eta_*,k) & = & g(\eta_*,k)\Phi(0,k),
\end{eqnarray}
where $f(\eta_*,k)$ and $g(\eta_*,k)$ are the transfer functions for
the respective quantities.

In Paper~I, using these transfer functions,
we derived a formula that relates the initial
curvature perturbation spectrum and the CMB angular correlation function
in a flat CDM universe under the
thin LSS approximation:
\begin{equation}
\tilde C(r)\equiv
3rC(r)+r^2C^{\prime}(r)=\frac{1}{2\pi^2}\int^{\infty}_{0} dk \,
P(k)\left\{f^2(k)k^2 r \cos kr +\lkk 2f^2(k)+g^2(k)\rkk k \sin kr \right\},
\label{cr+crd}
\end{equation}
where $P(k)=\langle |\Phi(0,k)|^2\rangle$
is the initial spectrum, $C(r)$ is the CMB
angular correlation function, and $r$ is the spatial distance between
two points on the LSS  sustained by an angle $\theta$.  Here
$C(r)$ is related with $\Theta_{\ell}(\eta_0,k)$ via the angular power
spectrum $C_{\ell}$ in the following manner.
\beq
C(r) = \sum_{\ell} \frac{2\ell+1}{4\pi}C_{\ell} P_{\ell}(\cos \theta),
~~~~~~~~~
\frac{2l+1}{4\pi}C_{\ell}=\frac{1}{2\pi^2} \int^{\infty}_{0} dk \, k^2
\frac{|\Theta_{\ell}(\eta_0,k)|^2}{2\ell+1}\,. \label{correlation}
\eeq

If we integrate by parts the right-hand side of Eq.~(\ref{cr+crd})
and use the Fourier-sine formula, we obtain
a first-order differential equation for $P(k)$,
\begin{equation}
-f^2(k) k^2 P^{\prime}(k)+\lkk-f(k)f^{\prime}(k) k+g^2(k)\rkk
kP(k)=4 \pi \int^{\infty}_{0}
\tilde{C}^{\thin}(r) \sin kr dr\, ,
\label{eqdif}
\end{equation}
where $\tilde{C}^{\thin}(r)$ should be calculated using
$\Theta_{\ell}^{\thin}(\eta_0,k)$ in (\ref{correlation}).
Because of the oscillatory nature of the transfer functions,
there appear zeros of $f(k)$ where Eq.~(\ref{eqdif}) becomes
singular. However, the values of $P(k)$ at these singularities
are readily known as
\beq
 P(k_i)=\frac{4 \pi}{k_i g^2(k_i)}\int^{\infty}_{0}
\tilde{C}^{\thin}(r) \sin k_ir dr\, ,
\eeq
where $k_i$ is the $i$-th zero point of $f(k)$.
Then the spectrum can be obtained easily and accurately
despite the presence of the singularities,
since we may solve it as a boundary value problem
between singular points rather than an initial value problem.

The thin LSS assumption, however, is not quite realistic
and $\Theta_{\ell}^{\thin}(\eta_0,k)$ deviates from the exact one
significantly.
The thickness of the LSS must be taken into account
for actual applications.
In this paper, we present a new method, which is basically
the same as the previous one but takes
account of the thickness of the LSS, and hence can
be applicable to realistic cases.

We first improve the transfer functions in the formula (\ref{eqdif})
by including the effect of the thickness of the LSS.
For this purpose, we use the following approximate expression
instead of (\ref{thin}).
\begin{eqnarray}
 \Theta_{\ell}^{\app}(\eta_0,k) & = &
(2\ell+1)\left[\int^{\eta_0}_{0} (\Theta_0+\Psi)(\eta,k){\cal V}(\eta) d\eta
\right]j_{\ell}(kd)
\nonumber\\
&&+(2\ell+1)\left[\int^{\eta_0}_{0}
\Theta_{1}(\eta,k){\cal V}(\eta)d\eta \right] j^{\prime}_{\ell}(kd)
\nonumber\\
& &+(2\ell+1)
\left[\int_0^{\eta_0}
\lmk\dot{\Psi}(\eta,k)
-\dot{\Phi}(\eta,k)\rmk e^{-\tau(\eta)}d\eta\right]
j_{\ell}(kd)\, ,
\label{thetal}
\end{eqnarray}
The first two terms of the above formula partially take account of
the thickness of the LSS,
which would be exact if the spherical Bessel functions
would not oscillate but remain constant over the thickness of the LSS,
which is the case for low wavenumber modes.
Note that the ISW term, which was ignored in Paper~I, is now included in
the third term. It takes account of the early ISW effect, since the
matter-radiation equality time is fairly close to the decoupling epoch.
Although the width of the time interval during which the integrand of
the third term is non-negligible is somewhat larger than the width
of the LSS, the early ISW effect happens to be described
by this approximation fairly accurately.
In fact, Eq.~(\ref{thetal}) turns out to be a very good approximation
for $C(r)$ or $C_\ell$
as is seen in Fig.~\ref{cl1}, in which an
angular power spectrum based on
our approximate formula is compared with the fully time-integrated one.
Note also that Eq.~(\ref{thetal}) is applicable to flat
$\Lambda$CDM models as well, if we neglect the late ISW effect which is
 important only for low-multipoles. In this case, we should not
 integrate the ISW term until $\eta_0$, otherwise it causes serious error
 since the late ISW effect is incorrectly included.

Defining new transfer functions $\fbar(k)$ and $\gbar(k)$ to express
(\ref{thetal}) as
\beq
\Theta_{\ell}^{\app}(\eta_0,k)=(2\ell+1)\fbar(k)\Phi(0,k)j_{\ell}(kd)+
(2\ell+1)\gbar(k)\Phi(0,k) j^{\prime}_{\ell}(kd), \label{thetaap}
\eeq
and repeating the same procedure as in Paper~I,
we arrive at the following differential equation.
\beq
-\fbar^2(k) k^2 P^{\prime}(k)
+\lkk-2\fbar(k)\fbar^{\prime}(k) k+\gbar^2(k)\rkk
kP(k)=4 \pi \int^{\infty}_{0}
\tilde{C}^{\app}(r) \sin kr dr\, ,
\label{newdif}
\eeq
where the right-hand-side should now be calculated using (\ref{thetal})
in (\ref{correlation}).

The above differential equation can be solved in the same way as
(\ref{eqdif}) to obtain $P(k)$.  Unfortunately, however, if we used the
observed correlation function obtained by, say, MAP satellite \cite{MAP}
in the right-hand-side of (\ref{newdif}),
we would reach an incorrect primordial spectrum $P(k)$,
because, although  the approximate formula (\ref{thetal})
reproduces the locations of peaks and troughs
of the true spectrum quite well, the amplitude still deviates from
the one obtained by full numerical calculation \cite{cmbfast}
which is to be compared with the observed one.

%
%

\begin{figure}[h]
\begin{center}
\leavevmode
\epsfxsize=7.0cm
\epsfbox{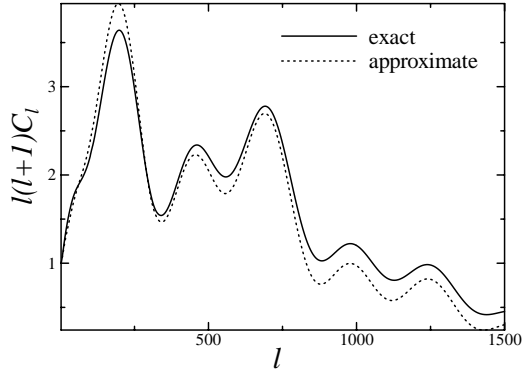}

\caption{The angular power spectra of CMB anisotropy in a spatially flat
universe
with $h=0.7$, $\Omega_b=0.03$, $\Omega_{\rm CDM}=0.97$
and scale-invariant adiabatic perturbation.
Solid line is based on fully time-integration,
while dashed line is the result of our approximate formula (\ref{thetal}).
The errors on large angular scales  with small $\ell$
are almost due to the approximation of the ISW term.
On small scales (large $\ell$), the effect of thickness of the LSS
becomes so significant that the approximation leads to large errors.
However, the qualitative features such as peak locations
are still reproduced relatively accurately.
}
\label{cl1}
\end{center}
\end{figure}

Fortunately, there exists a remedy for
errors caused by the formula (\ref{thetal}).  To show this,
let us consider the ratio of the exact to the
approximated angular power spectrum, which we define as
\begin{equation}
b_{\ell} \equiv \frac{C^{\rm exact}_{\ell}}{C^{\rm app}_{\ell}}\,.
\end{equation}
Of course, this depends on the shape of the initial spectrum.
However, the dependence turns out to be relatively weak as is
illustrated in Fig.~\ref{cl2}.
That is, for a fixed set of the cosmological parameters,
$b_\ell$ is found to be rather insensitive to
the variation of the initial spectrum.
This suggests the following procedure of reconstruction.
Consider we are given the observational data
($\equiv C^{\rm obs}_{\ell}$). Assuming
the cosmological parameters are known, we calculate
the correction factor $b_{\ell}$ for a fiducial initial spectrum
$P^{(0)}(k)$. Namely, we calculate the exact angular spectrum
$C^{(0)}_\ell$ and the approximate one $C^{{\rm app}(0)}_\ell$
using Eq.~(\ref{thetal}) from $P^{(0)}(k)$ and take their ratio
which we denote by $b^{(0)}_\ell$.
Then dividing $C^{\rm obs}_\ell$ by $b^{(0)}_\ell$,
we estimate $C^{\rm app}_{\ell}$.
Let us denote this estimated approximate angular spectrum
by $C^{{\rm app}(0)}_\ell$.
Then, we can insert it into
Eq.~(\ref{newdif}) to reconstruct the initial spectrum.
If $b_\ell$ were rigorously invariant, this procedure would
recover the true initial spectrum.
Due to small errors caused by the non-invariance of $b_\ell$, however,
the reconstructed spectrum, denoted by $P^{(1)}(k)$, will not be exactly
equal to the true primordial spectrum $P(k)$.  Nonetheless,
we can expect $P^{(1)}(k)$ to be fairly close to the true $P(k)$,
or at least better than the fiducial spectrum $P^{(0)}(k)$
which is a blind guess. Then, we can iterate this procedure
to improve the accuracy.

Schematically, this iterative procedure is described as
\begin{eqnarray}
P^{(n)}(k)\quad\to\quad
 b^{(n)}_\ell={C^{(n)}_\ell\over C^{{\rm app}(n)}_\ell}
\quad\to\quad
  C^{{\rm app}(n+1)}_\ell={C^{\rm obs}_\ell\over b^{(n)}_\ell}
\quad
\mathop{\longrightarrow}\limits_{\rm Eq.\,(\ref{newdif})}
\quad P^{(n+1)}(k)\,,
\end{eqnarray}
where the last step is the inversion procedure, and
$C^{{\rm app}(n)}_\ell$ is the approximate angular spectrum
for $P^{(n)}(k)$ using the formula (\ref{thetal}).
We repeat this procedure until we obtain a given degree
of convergence.

The validity of the above prescription
and the rate of convergence depend on the
degree of invariance of $b_\ell$.
In Fig.~\ref{cl2}, we plot $C_\ell$ and
$b_{\ell}$ for a scale-invariant and a peaked spectrum.
Although the difference of $C_\ell$ between the two spectra
is significant, the difference of $b_{\ell}$ is small,
namely, only a few percent.
The invariance of $b_\ell$, of course, will not hold
between entirely different initial spectra.
Recent observations, however, suggest that
the initial spectrum is almost scale-invariant
on very large scales. We therefore assume that
the scale-invariant spectrum is a good guess,
apart from possible features like peaks and dips
and/or some smooth variations of the power-law index
over a range of one or two orders of $k$.
\begin{figure}[h]
\begin{center}
\leavevmode
\epsfxsize=7.5cm
\epsfbox{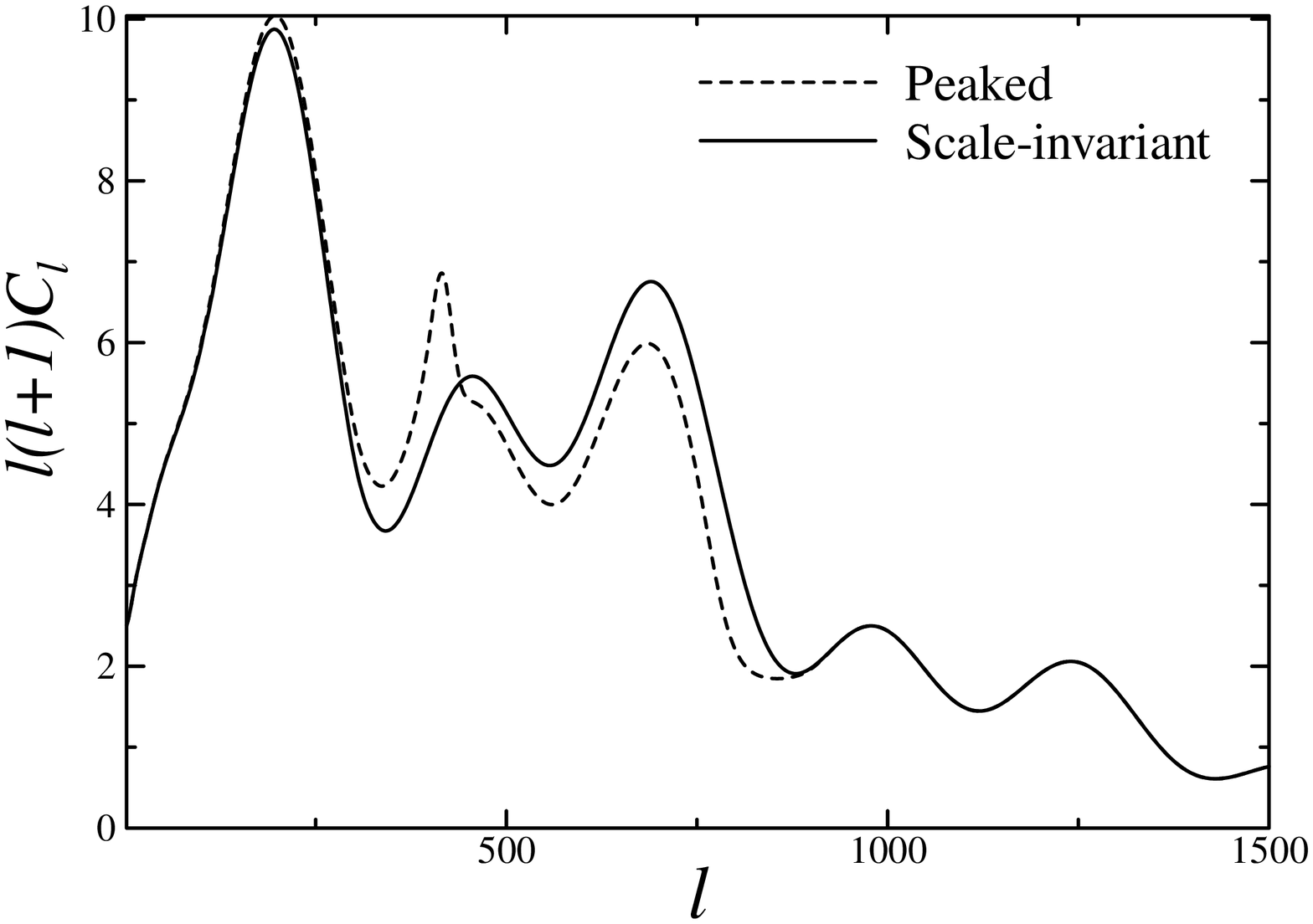}\qquad
\epsfxsize=7.3cm
\epsfbox{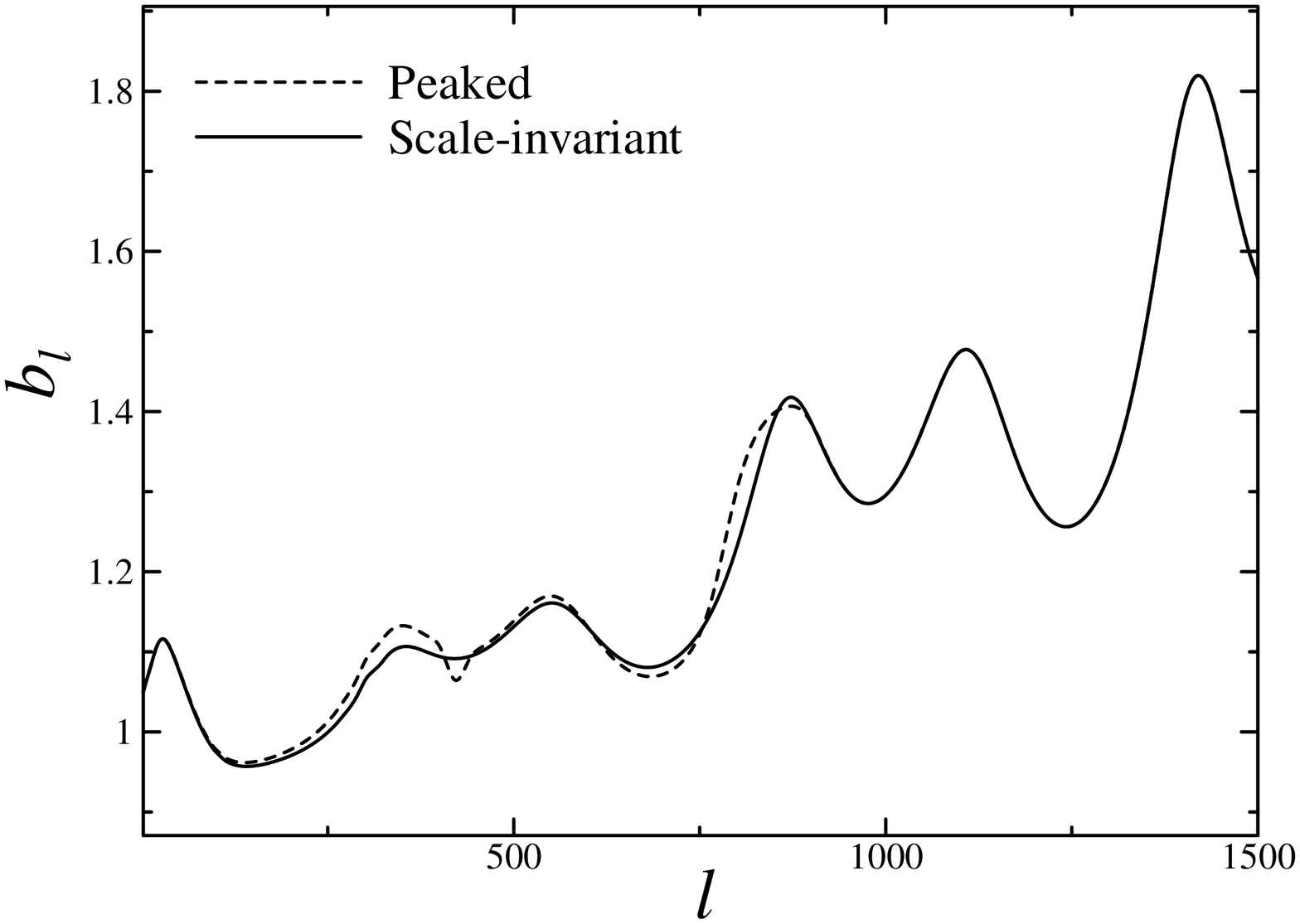}
\caption{The right panel shows the angular power spectra $C_\ell$
for scale-invariant (solid line) and peaked (dashed line)
initial spectra. The left panel shows the ratio
 $b_l=C^{\rm exact}_\ell/C^{\rm app}_\ell$ for the two initial spectra.}
\label{cl2}
\end{center}
\end{figure}

\section{Results}
In order to test the validity of our method,
we apply it to several shapes of initial spectra
in flat CDM models.
It may be noted that, although we focus on flat CDM models,
we method is equally applicable to flat $\Lambda$CDM models.
In fact, our models may be approximately regarded as
representing flat $\Lambda$CDM models
for the same value of $\Omega_{\rm CDM}h^2$\cite{HF}.
This is because the angular spectrum depends on $\Omega_{\rm CDM}$
only through the combination $\Omega_{\rm CDM}h^2$, except for
$C_\ell$ at small values of $\ell$ to which the late ISW effect
may contribute, but we do not use $C_\ell$ at $\ell\lesssim 30$
for the reconstruction.

For initial spectra having distinct features, a spectrum
obtained at the first round of inversion sometimes
contains regions of negative values or has extraordinary sharp
peaks or dips (with height or depth of more than
a factor of ten relative to the continuum)
due to errors in the estimation of $b_\ell$.
Since a real spectrum must be positive definite, we
make it a rule to cut off the negative part from the
spectrum and smooth out sharp peaks and dips to
mild ones, and use it for the next round of the inversion
procedure. Although there is no justification for this
prescription, we find that these peculiar features disappear
from the spectrum at the next round when they are spurious,
while they show up again when they are real.
We first assume the correct cosmological parameters
are known. Then the iteration converges by a few times,
although the rate of convergence depends on a spectrum.
We find the converged spectrum agrees well with the original
spectrum. In actual applications, we need to calculate
the corresponding angular spectrum again to compare
it with an observed spectrum $C^{\rm obs}_\ell$.

We plot the result for a spectrum with a peak and a dip
in Fig.~\ref{pkite}(a).
In this example, there appears a sharp dip at $kd \simeq 900$
at the first round of inversion. Assuming it is spurious,
we cut it off by hand and interpolate the spectrum smoothly
to proceed the iteration. In the next round,
it disappears completely.
As seen from the right panel of the figure, the iteration
converges remarkably fast. The small hump at $kd \simeq 400$
disappears without trace, while the dip near $kd \simeq 800$
grows deeper and approaches to the real shape.
The bottom panel shows the relative error at the fourth
round of iteration. Although the precision around the
singularities of Eq.~(\ref{newdif}) is relatively bad,
the error is still within 4\%. We also plot the result for a spectrum
which has smooth variations of the power-law index
in Fig.~\ref{pkite}(b). In this case, there
appears no spurious peak or dip through the iteration so that we can
easily recover the shape of the real spectrum without modification.

\begin{figure}[h]
\noindent
\begin{center}
\leavevmode
\epsfxsize=7.0cm
\epsfbox{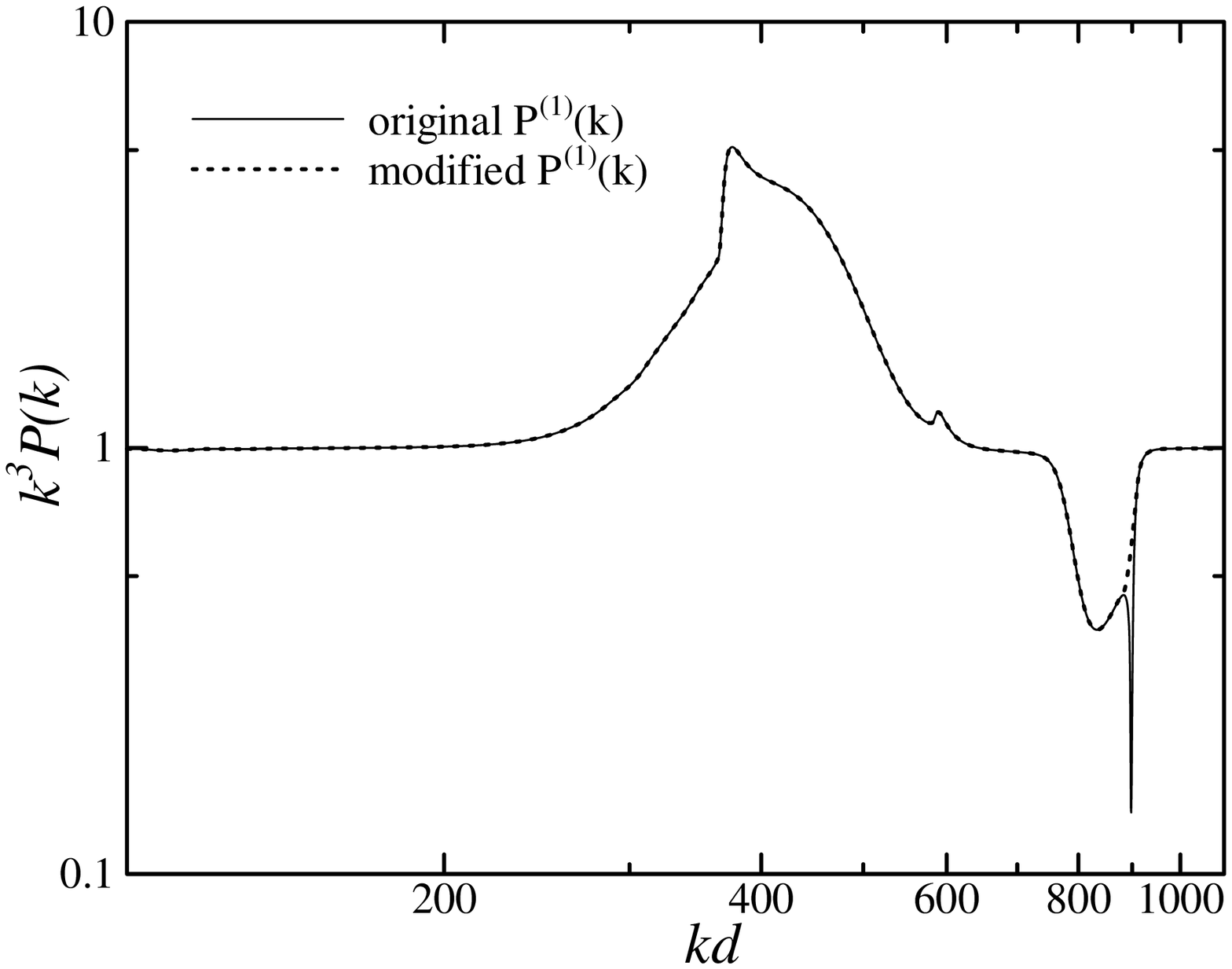}\qquad
\epsfxsize=7.0cm
\epsfbox{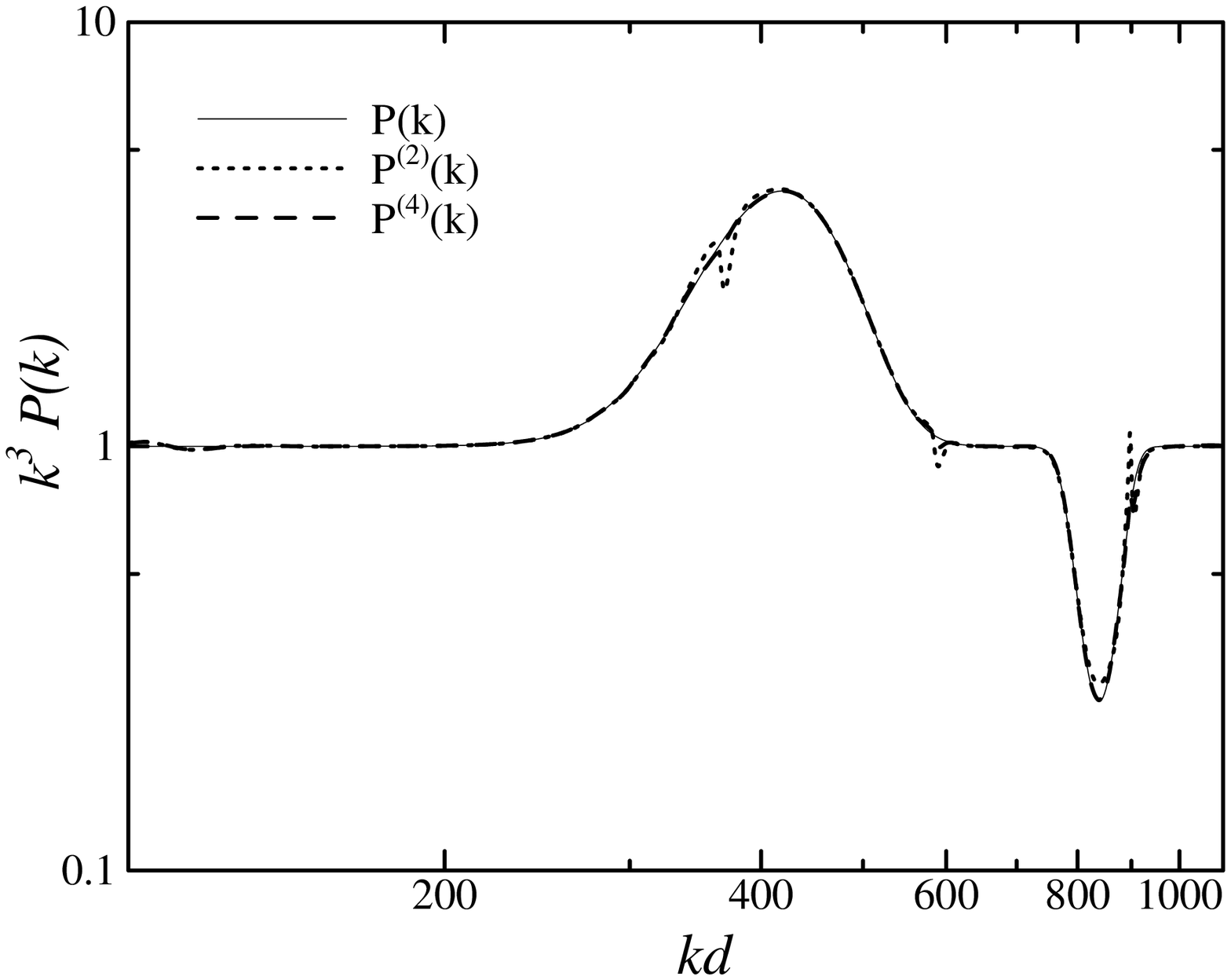}\\
\vspace{5mm}
\leavevmode
\epsfxsize=7.0cm
\epsfbox{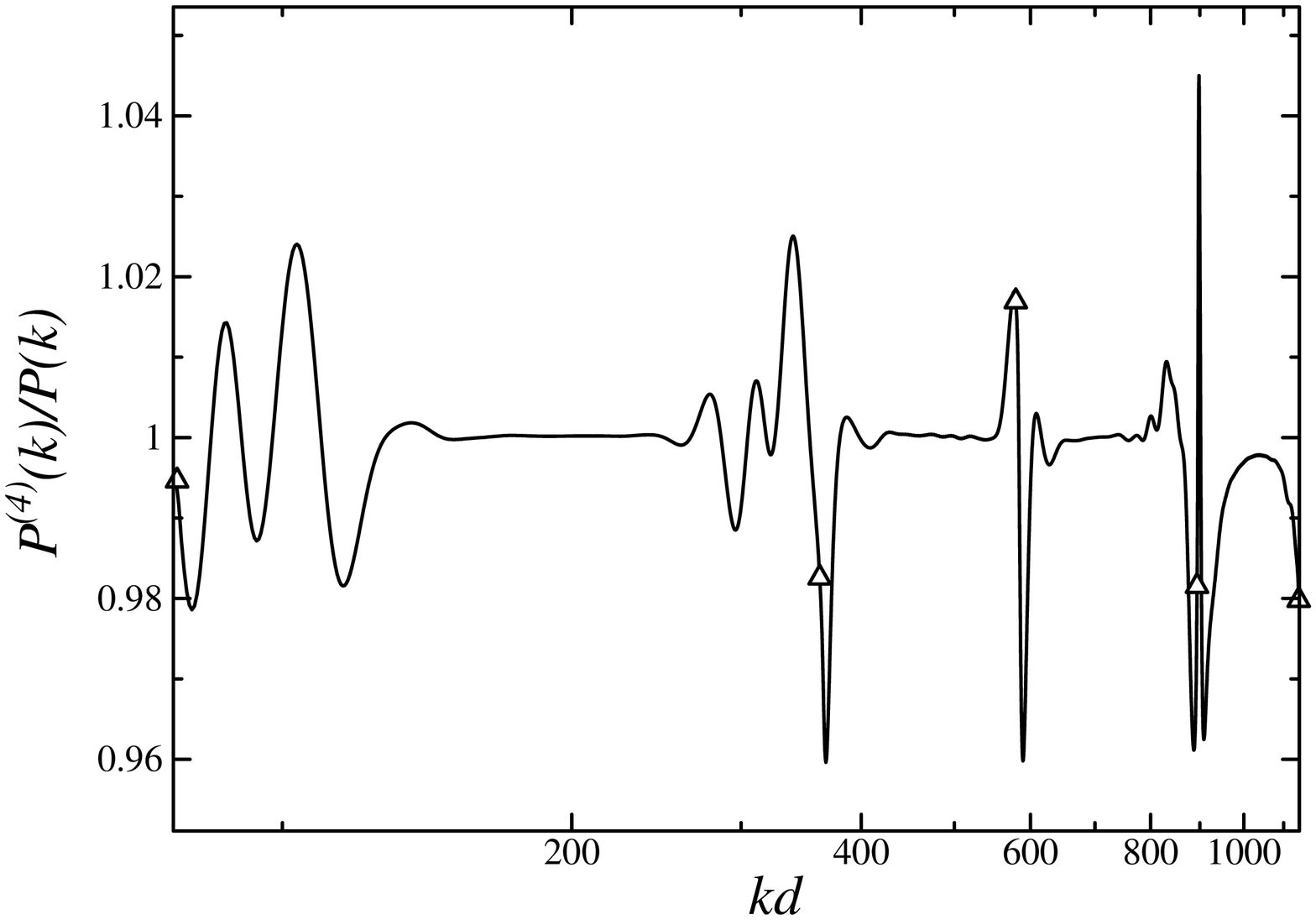}\\
Fig.~\ref{pkite}(a)
\end{center}
\vspace{5mm}
\begin{center}
\leavevmode
\epsfxsize=7.0cm
\epsfbox{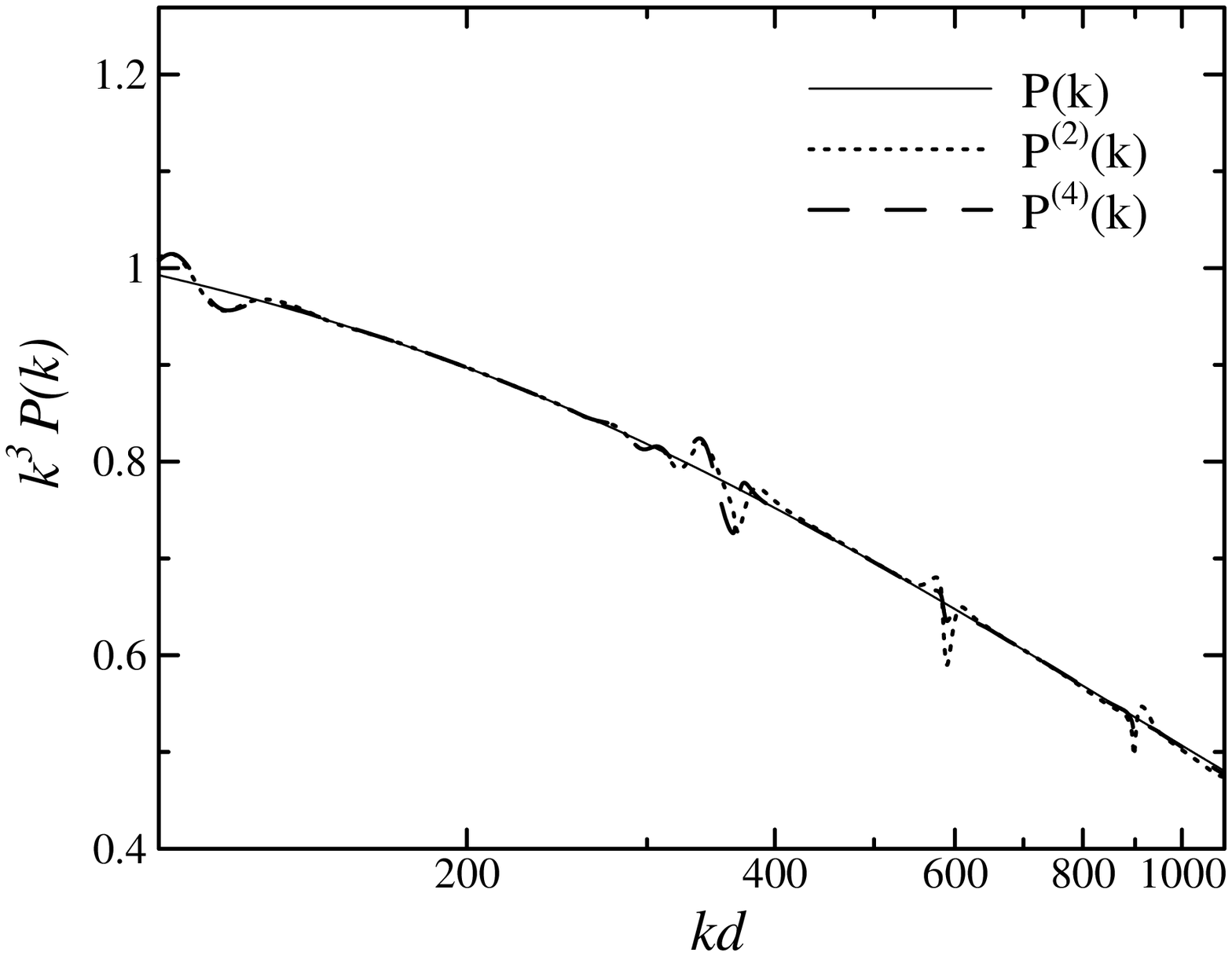}\\
Fig.~\ref{pkite}(b)
\vspace{5mm}

\caption{Test of our inversion method for a primordial
spectrum (a) with a peak and a dip superposed on a
scale-invariant spectrum and (b) with smoothly changing power-law index.
(a) The top left panel shows $P^{(1)}(k)$ which is obtained
from the first round of iteration. The solid curve is
the original $P^{(1)}(k)$ which happens to have a
very spiky dip at $kd\simeq 900$. We assume such a feature
is spurious and remove it by simply cutting that part
off and interpolating from the both sides, which is
plotted by the dotted curve.
The top right panel shows $P^{(2)}(k)$ and $P^{(4)}(k)$ together
with the original (true) spectrum $P(k)$.
The bottom panel shows $P^{(4)}(k)/P(k)$. The relative error is
negligible except for regions near the singularities (marked
by triangles) of the differential equation (\ref{newdif}).
Even near the singularities, the relative error is smaller
than $\sim 4\%$. (b) In this case, we do not need to modify
 $P^{(1)}(k)$. The result reproduces the variation of
the real spectrum well.
}
\label{pkite}
\end{center}
\end{figure}
In the above, we assumed that the resolution of observations is
limited to $\ell \leq \ell_{c}=1500$, which roughly corresponds to
$kd \leq k_cd=1500$ in the primordial power spectrum.
For our iterative procedure, however,
we must assume the shape of the spectrum at scales $k >k_c$,
because the contribution of modes at $k> k_c$ to the
angular spectrum at $\ell<\ell_c$ is non-negligible.
It becomes small enough only for $\ell\lesssim \ell_c/2$.
Since there is no principle to determine the spectrum at $k>k_c$,
we simply extrapolate the spectrum at $k<k_c$
to $k>k_c$ with a power-law form. This induces some errors
in the angular spectrum at $\ell\gtrsim \ell_c/2$.
In the case of the example shown in Fig.~\ref{pkite},
the original spectrum is assumed to be scale-invariant
at $k>k_c$ with the amplitude extrapolated from that at $k<k_c$.
Hence we were able to recover the angular spectrum over
the whole range up to $\ell=\ell_c$.
In an actual application of our method, however, we must decide not
to use $C^{\rm obs}_\ell$ at $\ell\gtrsim\ell_c/2$
as a criterion for the convergence.
\section{Constraining cosmological parameters}
In the previous section, we showed that the initial spectrum
can be determined with a high precision.
However we had to assume that the cosmological parameters,
which are usually estimated from the CMB data,
are exactly known by some other observations from the beginning.
Here we examine consequences of the use of incorrect cosmological
parameters in our procedure.

Naively, one might expect that our method is incapable of determining
the cosmological parameters, since a different choice of cosmological
parameters would simply give a spectrum that differs from the
true spectrum. This is true in the rigorous sense. However,
the very fact that the observed angular spectrum is used
to determine the cosmological parameters indicates that
there is not much freedom in varying cosmological parameters.
In fact, if we use parameters that differ significantly from
real values, very spiky features appear at the
singularities of Eq.~(\ref{newdif}) in the reconstructed
spectrum. Furthermore,
depending on the direction of deviation from the real values,
the reconstructed spectrum becomes negative in some
regions of $k$, usually near the singularities,
which never disappear by iteration.

As an example, Fig.~\ref{pkhab} shows the reconstructed spectra
for several values of $h$ and $\Omega_b$
when $C^{\rm obs}_{\ell}$ is given by
a scale-invariant spectrum with $h=0.7$ and $\Omega_b=0.03$
in the flat CDM universe.
As noted before, we may regard this case to describe a
flat $\Lambda$CDM models with $\Omega_{\rm CDM}h^2=0.7$ as well.
As for the $h$ dependence,
spiky peaks appear at the singularities for larger values of $h$,
while spiky dips appear for smaller values of $h$ and
the dips become too deep to render the spectrum negative for
$h \lesssim 0.65$.
Therefore, the positivity condition of the spectrum severely
constrains models with smaller values of $h$ and larger values of
$\Omega_b$,
while models with larger values of $h$ and smaller values of
$\Omega_b$ can be also
constrained unless there is a good reason to
believe that locations of the spikes and the singularities should
coincide.

\section{Conclusion}
We have presented a method to solve the inversion problem of
reconstructing the primordial curvature perturbation spectrum $P(k)$
solely from the CMB angular power spectrum $C_\ell$.
In Paper~I, we developed an inversion method by
deriving a first order differential equation for $P(k)$
with its source term determined by $C_\ell$,
but under the assumption of an infinitely thin LSS.
In this paper, we improved it significantly by fully incorporating both
the thickness of the LSS and the early ISW effect.
This was made possible because of an empirical fact that
the spectral ratio $b_\ell$ of the exact, full-numerically
calculated angular spectrum $C^{\rm exact}_{\ell }$ to
an approximate angular spectrum $C^{\rm app}_{\ell }$
using an analytic formula for the CMB multipoles (\ref{thetal}) is
relatively insensitive to variations of the shape of $P(k)$.
This fact allowed us to apply the inversion method developed
in Paper~I iteratively to reconstruct $P(k)$.
We have found the new method can recover the original spectrum
accurately, to within relative errors of 4\%, even in
cases when there are distinct features like peaks and
dips in the spectrum.

Our method, however, has a possible drawback that
the inversion procedure can be performed only when the values of
cosmological parameters are known. Therefore, we have also studied
the effect of using the cosmological parameters that are different
from their true values. We found that a small deviation
from the true values results in the appearance of peaks and dips
at the locations of singularities of the differential equation
for $P(k)$, which can be easily judged as spurious.
In particular, depending on the direction and the size
of deviations, the spurious dips can become so deep that
the positivity condition of $P(k)$ is violated there.
Thus, contrary to our concern about the inversion
method mentioned above, it may be regarded as an advantage
in the sense that the cosmological parameters can be constrained severely
with no regard to the shape of the initial spectrum.
This is in accordance with a widely accepted assertion that
the cosmological parameters can be determined from
the heights and locations of peaks in the observed
CMB spectrum. What is new is that we have not only justified
this assertion qualitatively but also provided a means to quantify
the level of its validity or limitations without any assumptions
on the form of the primordial spectrum.

Finally, let us point out a couple of issues to be explored
in future studies. First, we should investigate how observational errors
on $C_\ell$ would be reflected to the reconstructed spectrum of $P(k)$.
Second,
our method in the present form
can apply only to spatially flat universes. To make it more
general, an extension to cases of spatially curved universes
is necessary. In this respect, we may note the following.
Apart from the geometrical effect of curved space
that changes an angle sustaining a fixed distance on the LSS,
we may adopt the small-angle approximation which allows
us to use various flat space formulas. Hence, an extension
to non-flat universes seems feasible enough, if not straightforward.
Another issue is about the CMB polarization.
Our method assume that the CMB spectrum is dominated by
the scalar-type curvature perturbations. However, it
has been argued that the CMB spectrum may contain a non-negligible
contribution from tensor perturbations \cite{GW}.
To identify the tensor contribution to the CMB spectrum,
the CMB polarization spectrum plays a crucial role \cite{pol}.
Hence, to complete the inversion problem, it is necessary to develop
a formalism that utilizes both the temperature and polarization spectra
to reconstruct both the scalar and tensor perturbation
spectra simultaneously. These issues are currently under study.

\begin{figure}
\begin{center}
\leavevmode
\epsfxsize=7.0cm
\epsfbox{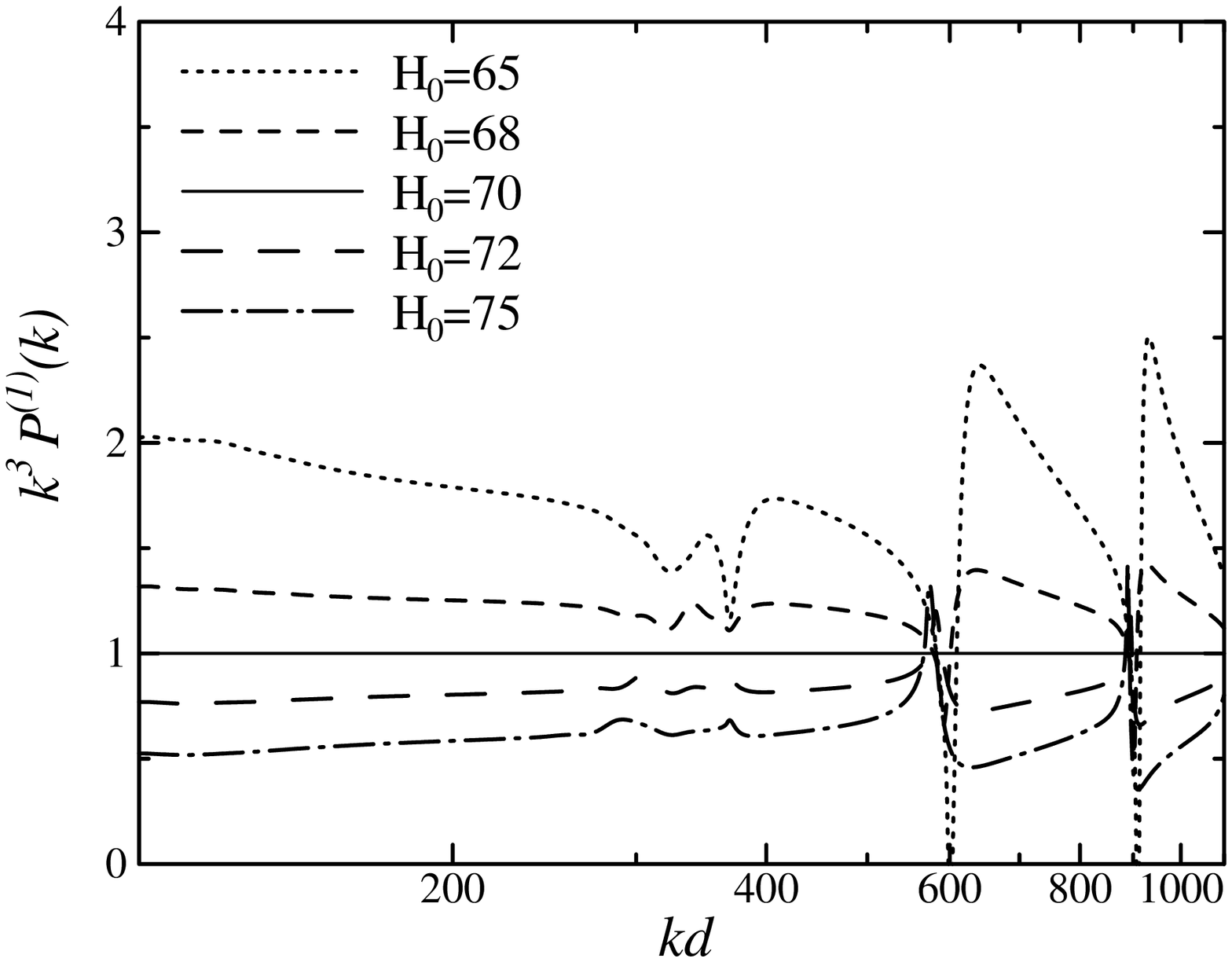}
\epsfxsize=7.0cm
\epsfbox{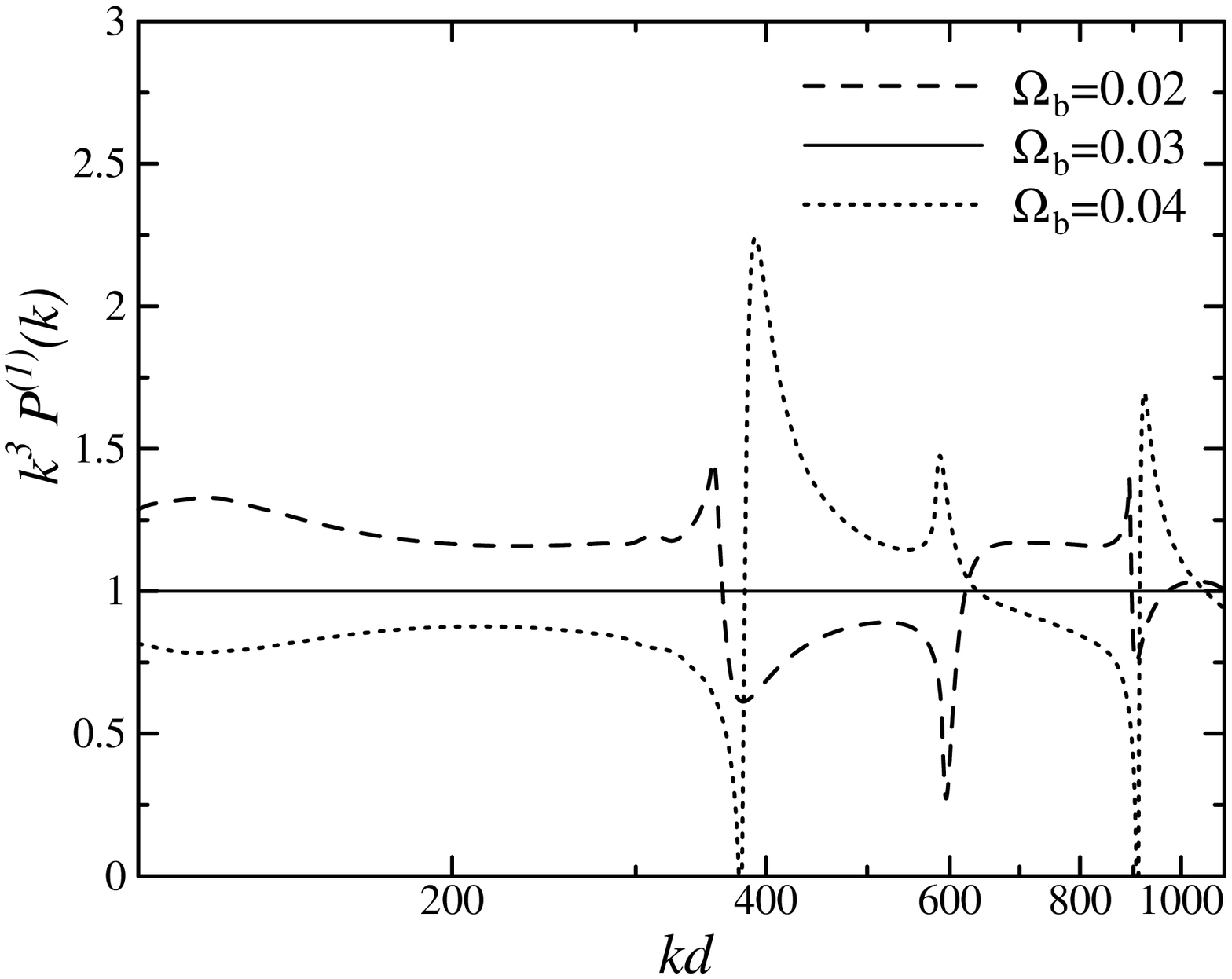}
\caption{The cosmological parameter dependence of the
reconstructed spectrum. The real spectrum is
 scale-invariant with $h=0.7$ and $\Omega_b=0.03$
in the flat CDM universe.
We plot the reconstructed spectra for
 $h=0.65$, $0.68$, $0.7$, $0.72$ and $0.75$, and
 $\Omega_b$=$0.02$, $0.03$ and $0.04$. }
\label{pkhab}
\end{center}
\end{figure}
\acknowledgements
This work was supported in part by JSPS Grant-in-Aid for
Scientific Research Nos.~12640269(MS) and 13640285(JY),
and by Monbu Kagakusho Grant-in-Aid for
Scientific Research (S) No.~14102004(MS).


\begin{thebibliography}{99}
\bibitem{FCDM}
A.~Balbi {\it et al.},
Astrophys.\ J.\  {\bf 545}, L1 (2000)
[Erratum-ibid.\  {\bf 558}, L145 (2001)]
[arXiv:astro-ph/0005124].
A.~H.~Jaffe {\it et al.}  [Boomerang Collaboration],
Phys.\ Rev.\ Lett.\  {\bf 86}, 3475 (2001)
[arXiv:astro-ph/0007333].
A.~E.~Lange {\it et al.}  [Boomerang Collaboration],
Phys.\ Rev.\ D {\bf 63}, 042001 (2001)
[arXiv:astro-ph/0005004].
C.~Pryke  {\it et al.}
Astrophys.\ J.\  {\bf 568}, 46 (2002)
[arXiv:astro-ph/0104490].
\bibitem{PER}
S. W. Hawking, Phys. Lett. {\bf 115B}, 295 (1982); A. A. Starobinsky, {\it
 ibid } {\bf 117B}, 175 (1982); A. H. Guth and S-Y. Pi,
 Phys. Rev. Lett. {\bf 49}, 1110 (1982).
\bibitem{MAP} http://map.gsfc.nasa.gov/
\bibitem{Planck} http://astro.estec.esa.nl/SA-general/Projects/Planck/

\bibitem{BSI}
A.~D.~Linde,
Phys.\ Lett.\ B {\bf 158}, 375 (1985).
L.~A.~Kofman and A.~D.~Linde,
Nucl.\ Phys.\ B {\bf 282}, 555 (1987).
J.~Silk and M.~S.~Turner,
Phys.\ Rev.\ D {\bf 35}, 419 (1987).
H.~M.~Hodges and G.~R.~Blumenthal,
Phys.\ Rev.\ D {\bf 42}, 3329 (1990).
J. Yokoyama and Y. Suto, Astrophys. J. {\bf 379}, 427 (1991);
M.~Sasaki and J.~Yokoyama,
Phys.\ Rev.\ D {\bf 44}, 970 (1991).
A.~A.~Starobinsky,
JETP Lett.\  {\bf 55}, 489 (1992);
[Pisma Zh.\ Eksp.\ Teor.\ Fiz.\  {\bf 55} (1992) 477];
J. Yokoyama, Astron. Astrophys. {\bf 318}, 673 (1997);
J. Yokoyama, Phys. Rev. D {\bf 58}, 083510 (1998);{\bf 59}, 107303 (1999);
S.~M.~Leach and A.~R.~Liddle,
Phys.\ Rev.\ D {\bf 63}, 043508 (2001)
[arXiv:astro-ph/0010082];
S.~M.~Leach, M.~Sasaki, D.~Wands and A.~R.~Liddle,
Phys.\ Rev.\ D {\bf 64}, 023512 (2001)
[arXiv:astro-ph/0101406].



\bibitem{SOU}
T. Souradeep et. al., astro-ph/9802262;
Y. Wang, D. N. Spergel and M. A. Strauss, astro-ph/9812291;
J.~Lesgourgues, S.~Prunet and D.~Polarski,
Mon.\ Not.\ Roy.\ Astron.\ Soc.\  {\bf 303}, 45 (1999)
[arXiv:astro-ph/9807020];
E.~Gawiser,
[arXiv:astro-ph/9807328];
Y.~Wang, D.~N.~Spergel and M.~A.~Strauss,
Astrophys.\ J.\  {\bf 510}, 20 (1999)
[arXiv:astro-ph/9802231];
S.~Hannestad,
Phys.\ Rev.\ D {\bf 63}, 043009 (2001)
[arXiv:astro-ph/0009296];
Y.~Wang and G.~Mathews,
Astrophys.\ J.\  {\bf 573}, 1 (2002)
[arXiv:astro-ph/0011351];

\bibitem{BER}
A. Berera and P. A. Martin, Inverse Problems {\bf 15}, 1393 (1999).

\bibitem{TEG}
M.~Tegmark and M.~Zaldarriaga,
Phys.\ Rev.\ D {\bf 66}, 103508 (2002)
[arXiv:astro-ph/0207047].
\bibitem{SW}
R.K.\ Sachs and A.M.\ Wolfe, Astrophys.\ J. {\bf 147}, 73 (1967).
\bibitem{MAT}
M.~Matsumiya, M.~Sasaki and J.~Yokoyama,
Phys.\ Rev.\ D {\bf 65}, 083007 (2002)
[arXiv:astro-ph/0111549].
\bibitem{HS}
W.~Hu and N.~Sugiyama,
Astrophys.\ J.\  {\bf 444}, 489 (1995).
\bibitem{KS}
H.\ Kodama and M.\ Sasaki, Prog.\ Theor.\ Phys.\ Suppl {\bf 78}, 1
	(1984)
\bibitem{cmbfast}
cmbfast; http://physics.nyu.edu/matiasz/CMBFAST/cmbfast.html

\bibitem{HF}
W.~Hu, M.~Fukugita, M.~Zaldarriaga and M.~Tegmark,
Astrophys.\ J.\  {\bf 549}, 669 (2001)
[arXiv:astro-ph/0006436].

\bibitem{GW}
A.A.\ Starobinsky, JETP Lett. {\bf 30}, 682 (1979);
V.A.\ Rubakov, M.V.\ Sazhin, and A.V.\ Veryaskin, Phys.\ Lett.\ {\bf
 115B}, 189 (1982).
L.F.\ Abbott and M.\ Wise, Nucl.\ Phys.\ {\bf B244}, 541 (1984);
R.\ Crittenden, J.R.\ Bond, R.L.\ Davis, G.\ Efstathiou, and P.J.\
Steinhardt, Phys.\ Rev.\ Lett. {\bf 71}, 324 (1993).


\bibitem{pol}
A.G.\ Polnarev, Soviet Astronomy, {\bf 29}, 607 (1985);
R.\ Crittenden, R.L.\ Davis, and P.J.\ Steinhardt,
Astrophys.J. {\bf 417}, L13-L16 (1993);


\end{thebibliography}
\end{document}